\documentclass[conference,a4paper]{IEEEtran}
\usepackage{cite}
\usepackage{amsmath,amssymb,amsfonts,mathrsfs,eufrak}
\usepackage{algorithmic}
\usepackage{textcomp}
\usepackage{subcaption}
\usepackage{soul,mathtools}
\usepackage{graphicx,lipsum}  
\graphicspath{{./figs/}}

\usepackage{xcolor}
\usepackage[left=1.61cm,right=1.61cm,top=1.81cm,bottom=4.35cm]{geometry} %for change size

\def\BibTeX{{\rm B\kern-.05em{\sc i\kern-.025em b}\kern-.08em
    T\kern-.1667em\lower.7ex\hbox{E}\kern-.125emX}}

\begin{document}

\title{Effect of Realistic Oscillator Phase Noise on the Performance of Cell-Free Massive MIMO Systems
}
\author{Igor Zhilin, Evgenii Vinogradov, Ian Akyildiz
\\\textsuperscript{}Autonomous Robotics Research Center, Technology Innovation Institute, UAE
\\\{Igor.Zhilin, Evgenii.Vinogradov, Ian.Akyildiz\}@tii.ae
}

\maketitle
%\vspace*{-0.8cm}
\begin{abstract}
As the demand for 6G technologies continues to grow, the radio access infrastructure is expected to become increasingly dense. Cell-free (CF) Massive MIMO systems provide remarkable flexibility by enabling coherent service to users through multiple Access Points (APs). This innovative paradigm necessitates precise and stable phase synchronization. This paper examines the standardized 5G New Radio (NR) framework, focusing on subcarrier spacing, OFDM symbol duration, and allocation, while investigating the impact of Phase Noise (PN) on the performance of scalable massive MIMO cell-free systems. Unlike existing studies that typically employ a simplified model of a free-running oscillator characterized by a Wiener process, we present a realistic phase noise model inspired by actual hardware, designed to accurately capture the Local Oscillator (LO) phase drift. Furthermore, our PN model extends its applicability beyond cell-free systems, making it relevant for any RF system operating within the sub-6 GHz band. This model provides a robust foundation for the practical design of cell-free systems, encompassing numerology and pilot allocation strategies. Our findings reveal that even cost-effective low-cost Local Oscillators can achieve sufficient stability, resulting in negligible degradation of uplink Spectral Efficiency (SE) within the standardized 5G Transmission Time Interval of 1 ms. These results affirm the viability of cell-free massive MIMO systems based on 5G standards and their potential integration into future 6G networks.
\end{abstract}

%\begin{IEEEkeywords}
%\end{IEEEkeywords}
\section{Introduction}
\bstctlcite{IEEEexample:BSTcontrol}
Cell-free (CF) massive multiple input-multiple output (mMIMO) systems are poised to play a critical role in the evolution of 6G networks \cite{9145564,9040264} as they address targets such as a 10-fold increase in user density, spectral efficiency (SE), and data rate compared to 5G. In CF mMIMO, user devices, commonly referred to as User Equipment (UE), are served by multiple Access Points (APs) that perform coherent joint transmission and reception \cite{7827017}. This innovative approach offers significantly enhanced SE compared to traditional architectures, such as Macro Base Stations or small cells \cite{bjornson2020scalable}, thereby promising a more efficient and effective network infrastructure for the forthcoming 6G era.

CF mMIMO has garnered considerable attention due to its remarkable performance demonstrated in theoretical studies conducted under ideal conditions \cite{bjornson2020scalable}. Building on this foundation, subsequent research has aimed to incorporate more realistic assumptions, including the use of non-ideal hardware, to facilitate practical implementation. These efforts can be classified into two primary categories: 1) hardware implementations or testbeds designed to validate and demonstrate the feasibility of CF mMIMO systems in real-world scenarios, and 2) analytical and simulation studies that adopt more realistic assumptions. Since coherent processing is the primary enabler of CF mMIMO, the phase synchronization between communication nodes (e.g., a UE and the APs serving it) is of paramount importance.

a) Testbeds: Wang et al. \cite{9183752} implemented a centralized cloud-based CF mMIMO system, where a CPU served 16 eight-antenna UEs with 16 eight-antenna APs. In another study \cite{9762865}, the KU Leuven CF Multiple-Input-Multiple-Output (MIMO) testbed was utilized to obtain channel state information (CSI) in a dense CF mMIMO deployment comprising 8 APs, each equipped with 8 antennas (64 antennas in total). In both studies, the effects of Hardware Impairments (HWI) were implicitly assessed due to the use of real equipment. However, both works relied on a common clock, and the impact of Phase Noise (PN) was not evaluated.

b) Theoretical Studies: Several investigations have explored the influence of Local Oscillator (LO) PN on single carrier \cite{9528977,9502552} and OFDM-based CF mMIMO systems \cite{Wu2023}. Inspired by \cite{4291833}, papers \cite{9528977,9502552,Wu2023} modeled PN by a discrete-time independent Wiener process, assuming independent innovations at each time instance. They observed significant performance degradation compared to ideal LO conditions, with drops of up to 58\% in terms of the 5\%-outage rate \cite{9502552} and a tenfold decrease in Spectral Efficiency within a 1~ms communication interval \cite{Wu2023}. The authors of \cite{9528977, Wu2023} proposed channel estimators that compensate for PN by estimating the effective channel. Specifically, they leveraged the assumed structure of PN in the channel estimator, subsequently using the PN-corrected channel estimates to calculate the Maximum Ratio (MR) and Minimum Mean-Square Error (MMSE) combining vectors, which are then employed to combine signals received from multiple APs into a single enhanced signal.

The PN model utilized in \cite{9528977,9502552,Wu2023} offers a reasonable approximation of actual LO performance. However, in \cite{9528977,9502552}, the LO phase drifts with a variance of 0.23 radians per OFDM symbol (71.4 us), whereas hardware manufacturers report phase noise that is an order of magnitude smaller for reasonably comparable frequency and time scales in affordable Software Defined Radios~\cite{USPR}. This discrepancy highlights the need for an improved PN model.

The contributions of this paper are as follows:
\begin{itemize}
    \item We provide a critical analysis of the impact of 3GPP time-frequency resource allocation on PN.
    \item We introduce a realistic, hardware-inspired phase noise model that accurately reproduces LO phase drift.
    \item We simulate a CF mMIMO system to evaluate the effect of realistic PN on the uplink SE of a 3GPP-compliant OFDM system.
\end{itemize}

To isolate and accurately assess the impact of PN, our study intentionally excludes other factors such as channel estimation errors and channel aging from our model.

\section{System Model}

In this section, we introduce the OFDM system model considering the effects of PN. Next, we adopt standard 3GPP settings to analyze their impact on the anticipated influence of PN. Finally, we present a scalable OFDM CF mMIMO system model, reflecting the realistic effects of PN.

Industry efforts in developing CF mMIMO systems are largely based on the 5G New Radio (NR) standards \cite{9647671,9979701}, which facilitates a smoother transition to practical implementation. In this paper, we follow a similar approach. To concentrate solely on communication performance, we assume that the procedures for Initial Access, Pilot Assignment, Cluster Formation, Channel Estimation, and Radio Resource Allocation are completed using techniques documented in the literature \cite{beerten2023cellfree,Girycki}.

\subsection{The OFDM System Model}
%\subsubsection{OFDM Transmission Model}

Consider an OFDM system comprising N subcarriers, with a subcarrier spacing of $\Delta f$ and a bandwidth defined as $B = 1/T_{\rm s} = N\Delta f$ where $T_{ s}$ represents the sampling period. The transmitted signal  $\{X_{i}\}_{i=0}^{N-1}$ is conveyed over $N$  subcarriers of an OFDM symbol, with an average per-symbol power of $p$. Utilizing an $N$-point Inverse Discrete Fourier Transform (IDFT), we can express the time-domain representation of the OFDM symbol transmitted by the UE as follows, where the index $n$ denotes the time-domain channel use (sample).
\begin{align}
\label{eq:idft}
x_{n} = {\frac{1}{\sqrt{N}}} {\sum_{i=0}^{N-1}} X_{i} e^{j 2 \pi i n/N},
\end{align}

\subsubsection{The Phase Noise Model}
Time-dependent random phase drifts, commonly known as PN, induce multiplicative distortions in the received signal. These distortions arise when the signal is multiplied by the output of the LO during the up- and down-conversion processes.

The total PN experienced between a user and an AP at the $n$-th time sample can be expressed as follows
\begin{equation}\label{eq:OFDM_PN}
    \theta_{n}=\phi_{n}+\varphi_{n},
\end{equation}
{where $\phi_n$ and $\varphi_n$ represent PN at the LOs of UE and AP, respectively.} 
The particular realizations of the $\phi_n$ and $\varphi_n$ depend on the model (see Sec. III).

We can express the time-domain PN realization during one received OFDM symbol as $\mathbf{p}^{\rm{time}} = [ e^{j \theta_{0}},\dots,e^{j\theta_{N-1}} ]^T \in \mathbb{C}^{N \times 1}$ while its frequency-domain counterpart is written as $\mathbf{p}^{\rm{freq}} = [ P_{0}, \dots, P_{N-1}]^T \in \mathbb{C}^{N \times 1}$ where the elements are obtained by DFT as $P_{i} = \frac{1}{N}	\sum_{n=0}^{N-1} e^{j \theta_{n}}e^{-j2\pi n i / N}$.

%Finally, $\mathbf{\Theta}_{mkn} = \mathrm{diag} \{e^{j \theta^l_{mkn}},\dots,e^{j\theta^L_{mkn}}\}$ is the corresponding PN matrix at the $n$-th channel use.
%The matrix is generic and can be easily transformed for a practical scenario where all antennas of an AP are connected to a single LO as $\mathbf{\Theta}_{mk,n} = e^{j \theta_{mkn}^1} \mathbf{I}_L$.

\subsubsection{The OFDM Signal Model with Phase Noise}
The received signal in the time domain ${\mathbf{y}^{\rm {time}}} \in \mathbb{C}^{N \times 1}$ is
\begin{align}
\label{eq:y_time}
\begin{split}
{\mathbf{y}^{\rm {time}}} 
&= 
\mathbf{p}^{\rm time} \circ ({\mathbf{x}^{\rm {time}}} \circledast {\mathbf{h}^{\rm {time}}}) + {\mathbf{z}^{\rm {time}}},
\end{split}
\end{align}
where $\mathbf{x}^{\rm {time}} \in \mathbb{C}^{N \times 1}$ the transmitted signal, $\mathbf{h}^{\rm {time}} \in \mathbb{C}^{N \times 1}$ the channel impulse response, and ${\mathbf{z}^{\rm {time}}} \in \mathbb{C}^{N \times 1}$ the additive white Gaussian noise~(AWGN) with i.i.d. ${\mathcal{CN}}\left(0, \sigma_{z}^2 \right)$ elements. It is easy to show that, the received signal in the frequency domain ${\mathbf{y}^{\rm {freq}}} \in \mathbb{C}^{N \times 1}$ is
\begin{align}
\label{eq:y_freq}
\begin{split}
{\mathbf{y}^{\rm {freq}}} 
&= 
{\mathbf{p}^{\rm {freq}}} \circledast ({\mathbf{x}^{\rm {freq}}}\circ {\mathbf{h}^{\rm {freq}}}) + {\mathbf{z}^{\rm {freq}}}
\end{split}
\end{align}
where $\mathbf{x}^{\rm {freq}} = [X_{0}, X_{1}, \cdots, X_{N-1}]^{\rm T}$, $\mathbf{h}^{\rm {freq}} = [H_{0}, H_{1}, \cdots, H_{N-1}]^{\rm T}$, $\mathbf{z}^{\rm {freq}} = [Z_{0}, Z_{1}, \cdots, Z_{N-1}]^{\rm T}\in \mathbb{C}^{N \times 1}$ are the transmit symbol, channel frequency response, and noise, respectively, in the frequency domain.

%The PN effects 
%********????**
%Y_{i}
%****
{The PN affected frequency domain received signal $Y$ for each subcarrier $i \in \{ 0,1, \cdots, N-1 \}$} %eqref{eq:y_freq} 
are expressed as \cite{9582730}
\begin{align}
\label{eq:y_freq_CPE_ICI}
Y_{i} = \underbrace{P_{0}}_{\rm CPE}H_{i}X_{i} 
+ \underbrace{{\sum_{\ell=0, \ell \neq i}^{N-1}} P_{(i-\ell)_{N}}H_{\ell}X_{\ell}}_{\rm ICI}
+ Z_{i},
\end{align}
%
%
%******
%***EXPLAIN all these notations in this equation*****
%*****
%
%
{where $(\cdot)_{N}$ denotes the modulo-$N$ operation. 

The Common Phase Error (CPE) $P_0$ represents an identical phase rotation across all subcarriers within an OFDM symbol, while the additive noise $P_\ell$  (which is not always Gaussian) constitutes the Inter-Carrier Interference (ICI) between carriers $i$ a and $\ell$. It is important to note that for a given OFDM symbol, the CPE can be estimated and compensated for during the channel estimation process.
% \cite{ZAIDI2018159}.

\subsubsection{The Channel Model}\label{Sec:Channel_OFDM}

We consider channel segmented into Coherence Blocks (CBs) characterized by a coherence time $T_{\rm c}$ and a coherence bandwidth $B_{\rm c}$. Within the coherence time, 
$N_{\rm ct} \triangleq  \lfloor {T_{\rm c}}/{T_{\rm OFDM}}\rfloor$
OFDM symbols, each with a duration of $T_{\rm OFDM}$, are transmitted. Similarly, we can calculate the number of subcarriers within the coherence bandwidth as $N_{\rm cb} \triangleq \lfloor{B_{\rm c}}/{\Delta f}\rfloor$. Each CB encompasses $N_{\rm ct}$ and $N_{\rm cb}$ successive OFDM symbols and subcarriers, during which the channel remains time-invariant and frequency flat. For our analysis, we will focus on one CB.

% \subsection{Practical Assumptions: Effect on the OFDM System Model with PN}\label{sec:assumptions}

\textbf{Numerology:} 
For Frequency Range 1 (FR1: sub-6 GHz), the predominant 5G NR numerology is $\mu=0$. Consequently, the smallest time-frequency resource unit allocated to a user corresponds to a CB spanning a bandwidth of 
$B_{\rm c}$=180~kHz
($N_{\rm cb}=12$ subcarriers with $\Delta f=15$~kHz spacing} and a coherence time of $T_{\rm c}=1$~ms (comprising $N_{\rm ct}=14$~OFDM symbols, each with a duration of $T_{\rm OFDM}=$71.4~$ \mu$s each). The total number of coherent channel uses is $\tau=N_{\rm ct}N_{\rm cb}$.

\subsubsection{Resource Element Mapping and Channel Estimation}

\begin{figure}
    \centering
    \includegraphics[width=1\linewidth]{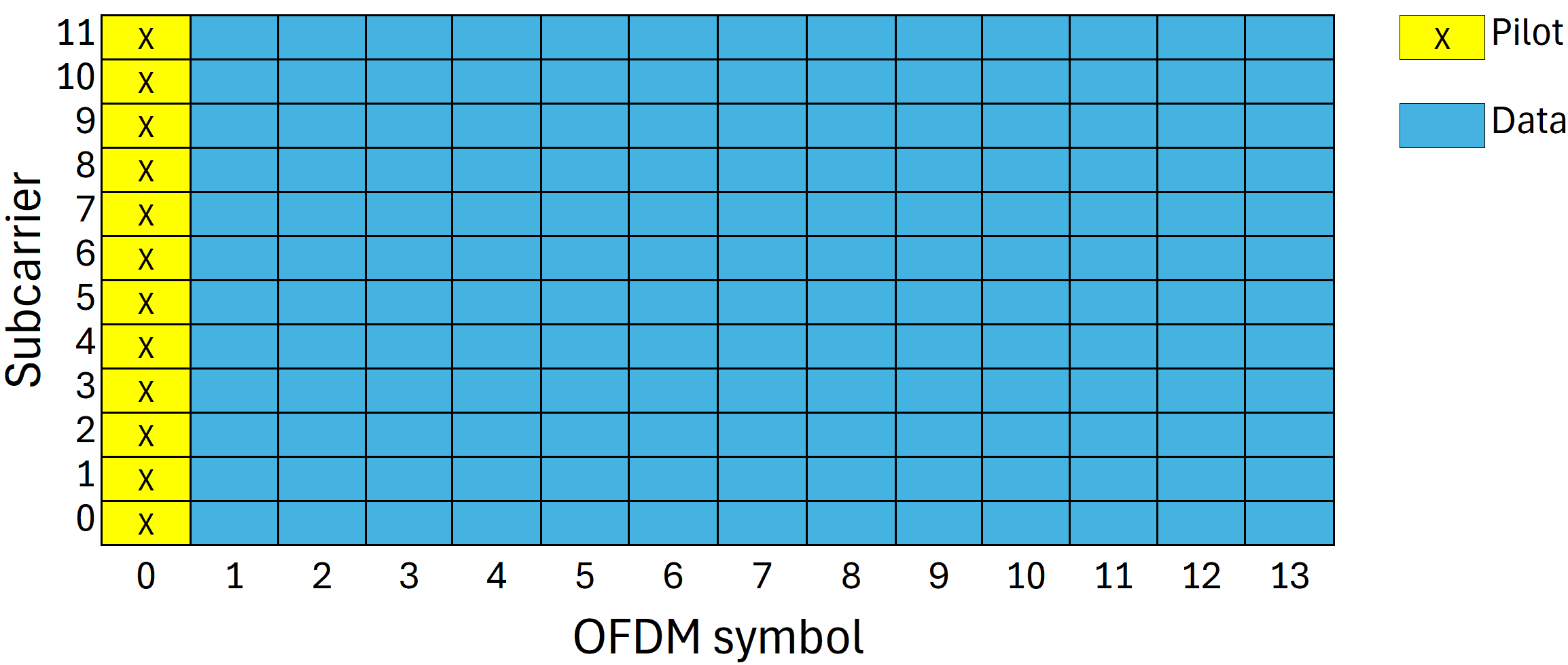}
    \caption{Mapping of the Resource Elements within a coherence block.}
    \label{fig:RE_mapping}
    \vspace*{-0.6cm}
\end{figure}

In accordance with \cite{Girycki}, we assume a Transmission Time Interval (TTI) format where all OFDM symbols are utilized for uplink communication. To facilitate channel estimation and CPE compensation, at least one OFDM symbol must include pilot signals (Demodulation Reference Signal) as per 3GPP TS 38.211. While OFDM symbols 3 or 4 are most commonly employed, we can consider a scenario where pilots are transmitted at the very beginning of the TTI (see Fig.~\ref{fig:RE_mapping}. The coherent channel uses  $\tau = \tau_p$ and $\tau_c$ comprise channel uses for pilot and data, respectively.

%\textbf{Effect on the OFDM system model:} 

The pilot allocation depicted in Fig.~\ref{fig:RE_mapping} represents a worst-case scenario from the CPE perspective. The CPE is compensated only during the $\mathfrak{t}=0$ OFDM symbol, while subsequent symbols are affected by an increasing error. For an arbitrary OFDM symbol $\mathfrak{t}$, the received signal can be expressed as follows.
\begin{align}
\label{eq:y_freq_CPE}
Y_{i}^{\mathfrak{t}} = P_{0}^{\mathfrak{t}} H X_{i}^{\mathfrak{t}} + Z_{i}^{\mathfrak{t}},
\end{align}
where $P_{0}^{\mathfrak{t}}$ is the CPE for the OFDM symbol $\mathfrak{t}$ (note that $P_{0}^0=1$) and $H$ is constant over $N_{\rm ct}$ OFDM symbols and $N_{\rm cb}$ subcarriers. {Similarly, $X_{i}^{\mathfrak{t}}$ and $ Z_{i}^{\mathfrak{t}}$ respectively denote the transmit symbol and noise for the subcarrier $i$ of the OFDM symbol $\mathfrak{t}$.}

 Compared to \eqref{eq:y_freq_CPE_ICI}, this equation also accounts for the fact that the 5G subcarrier spacing is designed to mitigate the effects of ICI.

\subsection{The Scalable Cell-Free system}

The CF mMIMO system depicted in Fig.~\ref{fig:CF} represents a communication network comprising multiple APs strategically distributed across a geographical area. These APs collaborate to serve $K$ users through coherent joint transmission and reception, utilizing shared time-frequency resources \cite{7827017}. Each AP is connected to CPUs responsible for coordinating the network. The operations of precoding and combining can occur at individual APs, centrally at the CPU, or through a hybrid approach involving both APs and CPUs \cite{9762865}. In this paper, we focus on centralized processing.

%, where a CPU conducts channel estimation and data detection based on signals received from multiple APs. This centralized scheme is widely employed in both theoretical and experimental works, including the aforementioned CF research (although some of the studies also investigate distributed architectures).

\subsubsection{The CF Channel Model}

We consider a general scenario where APs are equipped with $L$ antennas, extending the model presented in Sec~\ref{Sec:Channel_OFDM}. Within a coherence block, the random channel vector between UE $k$ and AP $m$ is modeled as independent correlated Rayleigh fading, represented by 
$\mathbf{h}_{mk} \sim \mathcal{CN}(0, \mathbf{R}_{mk})$ 
where, the small-scale fading is characterized by a complex Gaussian distribution, and 
$\mathbf{R}_{mk} \in \mathbb{C}^{L\times L}$ denotes a deterministic positive semi-definite correlation matrix that captures the large-scale propagation effects, such as shadowing and path loss, with a fading coefficient defined as  $\beta_{mk} = tr (\mathbf{R}_{mk})/L$. Consistent with \cite{bjornson2020scalable,9528977}, we assume that the set of correlation matrices is known to the communication nodes.

\subsubsection{Scalable CF}

\begin{figure}[!t]
\centerline{\includegraphics[width=0.5\linewidth]{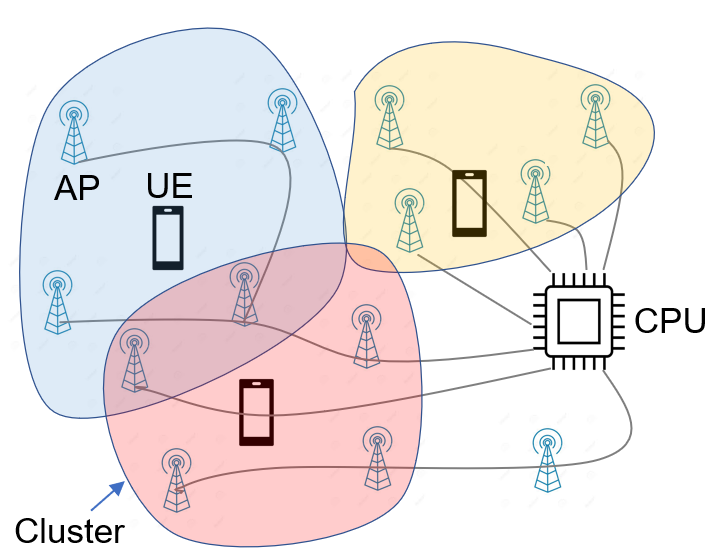}}
\caption{{CF massive MIMO system with user-centric clustering. }}
\label{fig:CF}
\vspace*{-0.6cm}
\end{figure}

 To promote network scalability, each AP is assigned to serve a specific set of UEs. 
 Bj{\"o}rnsson and Sanguinetti~\cite{bjornson2020scalable}  proposed the formation of clusters of APs that serve users in a user-centric manner (see Fig.~\ref{fig:CF}). In this configuration, AP $m$ exclusively serves a subset {$\mathcal{D}_m$} of $K$  UEs, ensuring network scalability, defined as follows:
 \begin{equation}\label{eq:clusters}
    \mathcal{D}_m = \{k: \quad \mathrm{tr} (\mathbf{D}_{mk})\geq 1, \quad k \in \{1,\dots,K\}\},
\end{equation}
%{where $\mathbf{D}_{mk} \in \mathbb{C}^{L \times L}$ 
where $\mathbf{D}_{mk} \in \mathbb{C}^{L \times L}$  is a diagonal matrix indicating the communication links between APs and UEs.

%***EXPLAIN...  the notation  (\mathbf{D}_{mk})\geq 1, \quad k \in \{1,\dots,K\}
%********

\subsubsection{The Phase Noise in CF}
%The phase noises at the $l$-th antenna LO belonging to $m$-th AP $\phi^l_m(t)$ and at the $k$-th user $\varphi_k(t)$ at the $t$-th channel use are stochastic processes. Similarly to \eqref{eq:OFDM_PN}, the total PN between a $k$-th user and $l$-th antenna of $m$-th AP can be expressed as
%\begin{equation}
%    \theta^l_{mkn}=\phi_{mn}^l+\varphi_{kn},
%\end{equation}
%where $n \in \{ 1,\cdots, \tau \}$ is the number of cherent channel use. Then, $\mathbf{\Theta}_{mk,n} = \mathrm{diag} \{e^{j \theta^l_{mkn}},\dots,e^{j\theta^L_{mkn}}\}$ is the corresponding PN matrix.

Given the absence of ICI, the phase noise is updated once per OFDM symbol, corresponding to an update of the CPE every $T_{\rm OFDM}$. In practical scenarios, where the $L$ antennas of the $m$-th AP are connected to a single LO, the phase noise matrix for the $k$-th UE can be expressed as $\mathbf{\Theta}_{mk}^{\mathfrak{t}} = e^{j \theta_{mk}^{\mathfrak{t}}} \mathbf{I}_L$
where $\theta_{mk}^{\mathfrak{t}}$ is derived as indicated in \eqref{eq:OFDM_PN} for AP $m$ and UE $k$. It is important to note that $\mathbf{\Theta}_{mk}^0 = \mathbf{I}_L$  as the CPE for $\mathfrak{t}=0$ is included in the channel estimation.

%The models of the phase noise are discussed in details in Sec.~III.

\subsubsection{The Received Signal Model}

 In this section, we derive the achievable SE for a practical scalable CF mMIMO system with phase noise. As previously mentioned, we consider centralized processing, where all received signals from the APs are forwarded to the CPU. The CPU then performs channel estimation and data detection. To isolate the negative effects of channel estimation errors from those introduced by phase noise, we assume perfect channel estimation. In the uplink scenario, the $M$ APs transmit the received signals  $ \{\mathbf{y}_{m,i}: m=1, \cdots, M \} $ to the CPU. 
 
 By setting $ W=ML $, the uplink signal received on the $i$-th subcarrier of the given OFDM symbol can be expressed as: 
 \begin{equation}
    \mathbf{y}_{i}=\sum_{k=1}^{K}\mathbf{\Theta}_{k} \mathbf{h}_{k} X_{k,i} + \mathbf{n},\label{ULTrans}
\end{equation}
where $X_{k,i} \in \mathbb{C} $  represents the signal from UE $k$ with power $ p_{k} $, and  $ \mathbf{y}_{i}=[\mathbf{y}_{1,n}^{T}\cdots \mathbf{y}_{M,n}^{T} ]^{T} \in \mathbb{C}^{W \times 1}$ is a block vector. The concatenated channel vector from all APs is denoted as 
$ \mathbf{h}_{k}=[ \mathbf{h}_{k1}^{T}\cdots \mathbf{h}_{kM}^{T} ]^{T} \sim \mathcal{CN} (\mathbf{0}, \mathbf{R}_k) $  is the phase noise matrix, including the CPE for this OFDM symbol. 
 
 The block diagonal spatial correlation matrix $ \mathbf{R}_{k}=\mathrm{diag} (\mathbf{R}_{k1}, \cdots,\mathbf{R}_{kM}) \in \mathbb{C}^{W \times W} $ is obtained under the assumption that the channel vectors from different APs are independently distributed. Finally, the thermal noise vector 
  $\mathbf{n} \sim \mathcal{CN}(0,\sigma^2_W\mathbf{I}_W)$ is spatially and temporally independent, with a variance of $\sigma_W$. We define the effective channel for the $t$-th OFDM symbol (including the CPE) as $\Tilde{\mathbf{h}}^{\mathfrak{t}}=\mathbf{\Theta}^{\mathfrak{t}}\mathbf{h}$.

As we stated before, although all APs receive the signals from all UEs, only a subset of the APs contributes to signal detection. Thus, the network estimates of {the signal $ X_{k,i} $ transmitted by user $k$ over subcarrier $i$ can be written as}
%
%
%****WHAT ARE THE $ X_{k,i} $*** WRITE OUT ***
%
%
\begin{equation}
\begin{split}
\hat{X}_{k,i}& = \sum_{m=1}^{M} \mathbf{v}_{mk}^{H}\mathbf{D}_{mk}\mathbf{y}_{m,i} \\ &\!=\!\underbrace{\mathbf{v}_{k}^{H}\mathbf{D}_{k}\Tilde{\mathbf{h}}_{k}X_{k}}_{Desired~signal}\!+\!\!\!\underbrace{\sum_{\ell=1,\ell\neq k}^{K}\!\!\!\mathbf{v}_{k}^{H}\mathbf{D}_{k}\Tilde{\mathbf{h}}_{\ell}X_{\ell}}_{Multi-user~interference}\!+\mathbf{v}_{k}^{H}\mathbf{D}_{k}\mathbf{n} \label{estimatedSignal} 
\end{split}
\end{equation}
with $ \mathbf{v}_{k}=\left [\mathbf{v}_{1}^{T}\cdots \mathbf{v}_{M}^{T} \right ]^{T} \in \mathbb{C}^{W \times 1}$ being the collective combining vector and $ \mathbf{D}_{k} =\mathrm{diag}\left (\ \mathbf{D}_{1k}, \ldots, \mathbf{D}_{Mk}\right )\in\mathbb{C}^{W \times W} $ being a block diagonal matrix while the bottom version of the equation is obtained from the top one in terms of collective vectors. 

%\zh{I think, in \eqref{estimatedSignal} and \eqref{eq:SINR}, i should replace $\mathbf{h}$ with an effective channel including the phase noise $\Tilde{\mathbf{h}}=\mathbf{\Theta}\mathbf{h}$, right?}

In this paper, we use MR and MMSE given as
\begin{align}
    \mathbf{v}_{k}^{\text{MR}} &= \mathbf{D}_k \hat{\mathbf{h}}_{k} 
    \\
    \mathbf{v}_{k}^{\text{MMSE}} 
    &= p_k \Big(\sum\limits_{\ell=1}^{K} p_\ell \hat{\mathbf{H}}_{\ell}^{D}  + \mathbf{Z}_{k} \Big)^{\dagger} \mathbf{D}_k \hat{\mathbf{h}}_{k},
    \label{eq:v_k_MMSE}
\end{align}
where $p_k$ is the transmit power of user $k$, $\hat{\mathbf{H}}_{\ell}^{D} =\mathbf{D}_k \hat{\mathbf{h}}_{\ell}
\hat{\mathbf{h}}_{\ell}^{H,} \mathbf{D}_k$ and $\mathbf{Z}_{k}=\mathbf{D}_k\big(\sum_{\ell=1}^K p_i \mathbf{R}_{i}+\sigma_{\text{}}^2 \mathbf{I}_{L}\big) \mathbf{D}_k$. 

\begin{figure}[t!]
\centerline{\includegraphics[width=.8\columnwidth]{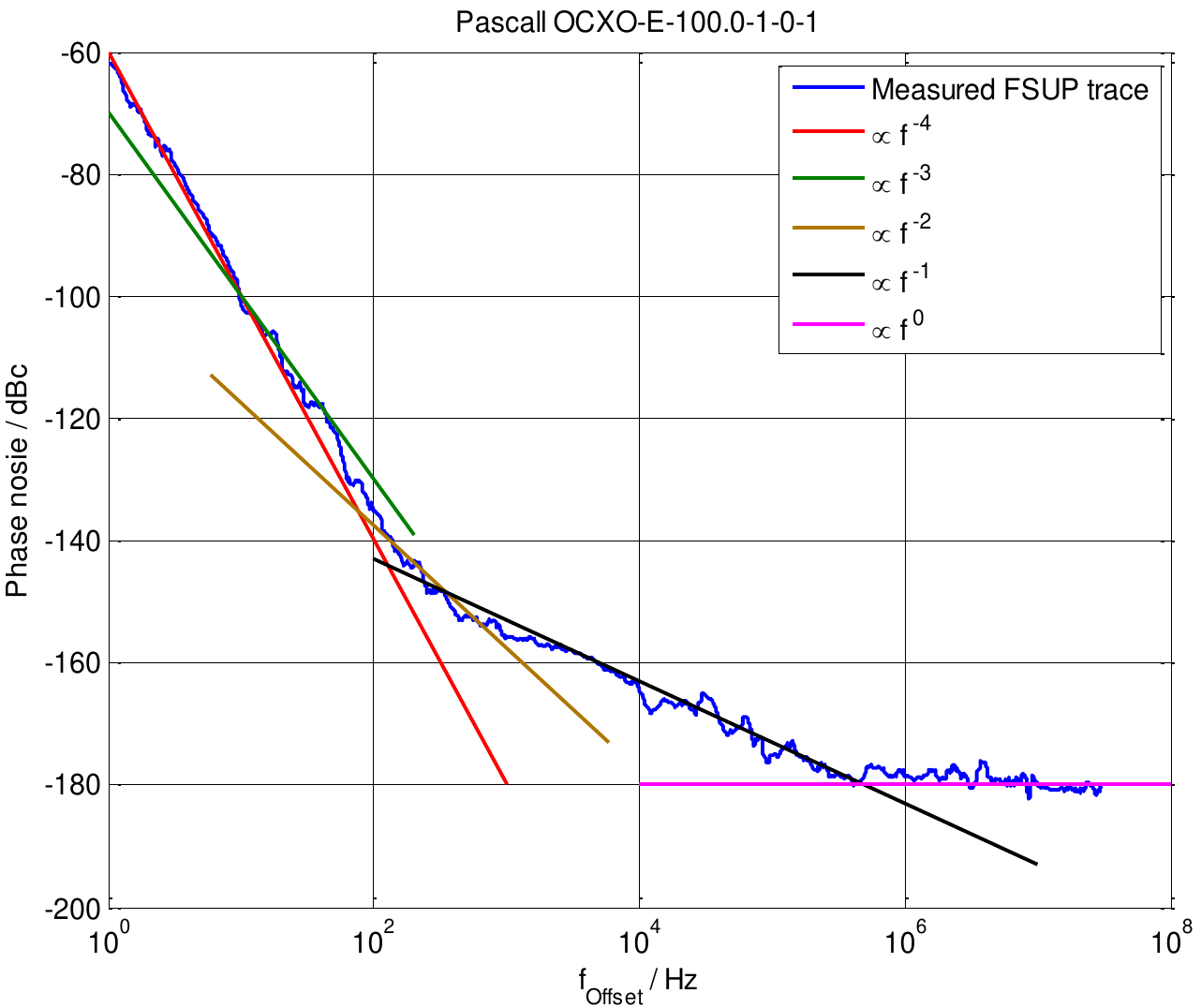}}
\caption{PSD of an ovenized crystal oscillator (OCXO) phase noise versus frequency offset with asymptots corresponding to different noise parts: phase white noise $~\sim f^{0}$, phase flicker noise $~\sim f^{-1}$, frequency white noise $~\sim f^{-2}$, frequency flicker noise $~\sim f^{-3}$, frequency random walk $~\sim f^{-4}$. Courtesy of Rohde\&Schwarz \cite{RSAN2015AllanVar}.}
\label{fig:pn_ocxo_w_asyptots}
\vspace*{-0.6cm}
\end{figure}

Finally, an achievable SE of UE $k$ for OFDM symbol $\mathfrak{t}$ is %given by
\begin{align}
    \text{SE}_{k}^{\mathfrak{t}} = \log_2(1+\text{SINR}_{k}^{\mathfrak{t}}),\label{eq:SE}
\end{align}
where Signal to Interference and Noise Ratio (SINR) is given by \eqref{eq:SINR}. The SE expression in~\eqref{eq:SE} can be computed numerically for any combiner  $\mathbf{v}_{k}$. Note that SE changes for each OFDM symbol due to the CPE.
\begin{figure*}[!h]
\begin{equation}
\text{SINR}_{k}^{\mathfrak{t}} = \frac{p_k \left|\mathbb{E}\left\{\mathbf{v}_{k}^{H} \mathbf{D}_k \Tilde{\mathbf{h}}_{k}^{\mathfrak{t}} \right\}\right|^2}
{\sum\nolimits_{\ell=1 }^{K} p_\ell  \mathbb{E}\Big\{\left|\mathbf{v}_{k}^{H  } \mathbf{D}_k \Tilde{\mathbf{h}}_{\ell}^{\mathfrak{t}} \right|^2\Big\} - p_k \left|\mathbb{E}\Big\{\mathbf{v}_{k}^{H} \mathbf{D}_k \Tilde{\mathbf{h}}_{k}^{\mathfrak{t}} \right\}\Big|^2 + \sigma^2\mathbb{E}\Big\{\left| \mathbf{D}_k \mathbf{v}_{k}^{H  } \right|^2\Big\}
}\label{eq:SINR}
\end{equation}
\hrule
\vspace*{-0.6cm}
\end{figure*}

\section{The Phase Noise Model}

% To have long-term stable and low phase noise references, the SDRs use high frequency voltage-controlled oscillators (VCO) synchronized to quartz crystal oscillators (XO) by phase-locked loops (PLLs). Depending on the particular device, this synchronization may include multiple stages PLL and, thus, multiple VCO and/or XO.

In this section, we begin by presenting a simplified PN model that is commonly referenced in the literature (Sec.~\ref{sec:simple_pn}). Next, we develop a more realistic PN model based on the practical implementation of Software Defined Radio (SDR) transceivers. In Sec.~\ref{sec:xopn}, we briefly discuss the contributions to PN introduced by various components of a LO. We conclude this section by presenting data from the actual devices we utilize for modeling in  Sec.~\ref{sec:sdrpn}.

\subsection{\label{sec:simple_pn}Simplified PN Model: Free Running Oscillator} 

As a reference, we employ a model in which the phase follows a Wiener process \cite{4291833}, described by $\phi_m(t) \sim \mathcal{N}(\phi_m(t-1),\sigma^2_\phi)$  and $\varphi_k(t) \sim \mathcal{N}(\varphi_k(t-1),\sigma^2_\varphi)$, where $t$ indicates the channel use number, while the variance of innovation is given by
\begin{equation}\label{eq:var}
    \sigma_x^2=4\pi^2 f^2_c c_x T_{\rm OFDM},
\end{equation}
where $x$ is to be replaced by $\phi$ or $\varphi$ while $f_c, c_x$ and $T_{\rm OFDM}$ are the carrier frequency, an oscillator-specific constant, and the symbol interval, respectively.

%\paragraph{Realistic PN model}

%Denoting $\Delta t=t_2-t_1$ where $t_1$ and $t_2$ are time instances when the channel is used (e.g., for coherent combining of signals and channel estimation, respectively), the phase drift can be modeled as a zero-mean Gaussian variable with a variance increasing with time:
%\begin{equation}
%    \Delta\phi \sim \mathcal{N} ( 0,\sigma^2_\phi\Delta t ).
%\end{equation}
%Following \cite{9528977}, the expectation of the complex exponential phase shift can be calculated as $\mathbb{E}\{e^{-j\Delta\phi}\} = e^{-\frac{\sigma^2_\phi}{2}\Delta t} $.

\subsection{Hardware-Inspired Phase Noise Model}

%To describe the PN of an LO, we must take into account all contributions of the crystal oscillator (XO), voltage-controlled oscillator (VCO), and the PN filtering done by the Phase-Locked Loops (PLLs). 

For accurate and realistic PN modeling, we implement a model of real-world devices. We consider SDR transceivers which use LOs for frequency up-/down conversion and sample I/Q at baseband. In our model, a LO is a voltage-controlled oscillator (VCO) which is locked to a reference crystal oscillator (XO) by a phase-locked loop (PLL).

This model allows us to precisely model the behavior of many common SDRs which are based on integrated transceiver chips (e.g., Analog Devices AD9361). We also apply this model to more complex SDRs which may have multiple PLL stages to filter the reference clock and generate the sampling and carrier frequency (e.g. Ettus USRP X300 series). In the latter case, we limit our consideration to the last VCO+PLL (the one that generates the actual carrier frequency) and to one XO.

%Below, we explain the PN power spectral density (PSD) figures we use as the input parameters of our model (sec.~\ref{sec:xopn}), then the way the PSDs of XO and VCO are combined by PLL (sec.~\ref{sec:pllpn}). Then in sec.~\ref{sec:sdrpn} we list the devices we consider and the PN figures we obtain for these devices. These PN figures are then used to initialize a PN random process generator implementation from Matlab Communications Toolbox to generate the $\mathbf{\Theta}_{mkn}$.

\subsection{\label{sec:xopn}The Phase Noise Model of a Local Oscillator}
% sources of PN
% how PN is described: SSB noise, jitter as time and as phase, maybe mention Allan variance

%In our model, we consider PNs of XOs and VCOs. 

To describe the PN of a standalone oscillator, the engineering community employs one-sided Power Spectral Density (PSD) denoted as $S(\delta f)$ where $\delta f$ represents the frequency offset from the carrier frequency. Depending on the time scale under consideration,  $\delta f$ 
can range from fractions of Hertz—corresponding to time intervals of one second or more—to hundreds of MHz, particularly when analyzing period-to-period jitter of a carrier.
Various sources of PN  dominate at different time scales: at high $\delta f$ white and flicker noise of phase are prevalent, while at slower time scales, white and flicker noise of frequency, as well as frequency random walk, become significant. An illustration of the contributions from these noise types is shown in Fig.~\ref{fig:pn_ocxo_w_asyptots}.

Crystal oscillators exhibit good stability over long time scales and moderate phase noise at high $\delta f$. However, they are not suitable for directly generating frequencies exceeding approximately $\sim 100$~MHz. When the frequency is multiplied to the carrier frequency $f_c$, the amplitude of phase noise increases proportionally with the carrier frequency:
\begin{equation}
    S_{f_c}(\delta f) = S_{\mathrm{XO}}(\delta f) \cdot f_c / f_{\mathrm{XO}}.
    \label{eq:pnmul}
\end{equation}

In contrast, VCOs can directly generate high frequencies and exhibit low phase noise at high $\delta f$, but their frequency stability is notably poor. To create a frequency source that combines both good stability and low noise, XOs and VCOs are often integrated using PLL.

%\subsubsection{\label{sec:pllpn}PLL}
 
\begin{figure}[]
\centerline{\includegraphics[width=.75\columnwidth]{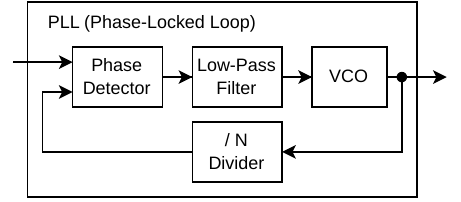}}
\caption{A single loop of PLL.}
\label{fig:PLL}
\vspace*{-0.6cm}
\end{figure}

By utilizing a PLL, it is possible to obtain a high-frequency oscillator that exhibits: i) low PN from the VCO at high-frequency offsets, and ii) low PN from the XO at low-frequency offsets, ensuring both long- and short-term stability.  Fig.~\ref{fig:PLL} illustrates a block diagram of a PLL. The phase detector receives two inputs: a reference frequency and a divided VCO frequency, which it uses to compare the phase difference. The resulting phase difference signal is then low-pass filtered to tune the VCO frequency. This low-pass filter effectively removes the high-frequency noise component from the XO. 

Consequently, at frequency offsets greater than the cut-off frequency $f_{\mathrm{PLL}}$, the PN is primarily determined by the VCO. Conversely, at frequency offsets below $f_{\mathrm{PLL}}$, the LO PN is influenced by the PN of the XO (amplified according to equation \eqref{eq:pnmul}) and the PN contribution from the phase detector. The total PN: %of the LO is
\begin{equation}
    S_{\mathrm{LO}}(\delta f) = S_{\mathrm{PLL+VCO}}(\delta f) + \frac{S_{\mathrm{XO}}(\delta f) \cdot f_c / f_{\mathrm{XO}} }{\sqrt{1 + (\delta f/f_{\mathrm{PLL}})^{2}}},
    \label{eq:pllpn}
\end{equation}
where %$f_c$ is the carrier frequency and 
$f_{\mathrm{XO}}$ is the original XO frequency.

For our simulations, we gather closed-loop performance data of the PLL combined with the VCO, denoted as $S_{\mathrm{PLL+VCO}}(\delta f)$, from the datasheets of specific integrated circuits (ICs). This data encapsulates the contributions from both VCO noise and phase detector noise.

%\subsubsection{\label{sec:chainpn}cascaded PLL}
% TODO: it would be nice to draw a chain of Osc:
% [Extrernal Sync as 1 PPS or 10 MHz] --<PLL1 @ Hz>-- [Internal XO (Crystall Osc) at 10..100 MHz] --<PLL2 @ kHz>-- [Clock distribution LO at 100's of MHz] --<PLL3 @ MHz>-- [RF LO producing 2 GHz]
% in simpler devices and in our model it is:
% [Internal XO (Crystall Osc) at 10..100 MHz] --<PLL @ MHz>-- [RF LO producing 2 GHz]

% \begin{figure}[!h]
% \centerline{\includegraphics[width=1.\columnwidth]{figs/PLL_chain.drawio.pdf}}
% \caption{A chain of PLLs}
% \label{fig:PLL_chain}
% \end{figure}

% Figure 2: A chain of PLLs and the resulting PN\\
% Trivial?\\
% Show components and combined
\begin{table*}
\centering
\begin{tabular}{| c | c | c | c | c | c | c | c | c | c |} 
\hline
$\delta f$ & 1~Hz & 10~Hz & 100~Hz & 1~kHz & 10~kHz & 100~kHz & 1~MHz & 10~MHz \\
\hline
USRP B200 & -26.0 & -52.0 & -76.7 & -93.9 & -101.8 & -102.7 & -113.1 & -138.1 \\ 
\hline
USRP 2954R & -22.0 & -62.0 & -89.9 & -103.7 & -101.0 & -106.0 & -137.4 & -156.0 \\
\hline
Instrumental & -76.5 & -90.9 & -100.1 & -106.5 & -107.8 & -108.0 & -134.9 & -154.9 \\
\hline
\end{tabular}
\caption{The calculated phase noise of the considered devices at different frequency offsets. Values at offsets from 10~Hz to 100~kHz are used for simulating the cell-free scenario with the considered 5G NR numerology.}
\label{table:pn}
\vspace*{-0.6cm}
\end{table*}
\subsection{\label{sec:sdrpn}Real Devices and Their Performance}

In this work, we focus on the stability of oscillators over time scales relevant to the 5G NR protocol, ranging from the duration of an OFDM symbol $T_{\mathrm{OFDM}} = 71.4$~$\mu$s to the duration of a frame $T_{\mathrm{c}} = 1$~ms. Therefore, our area of interest is defined by the frequency range 
\begin{equation}
    10\mathrm{~Hz} \ll 1 / T_{\mathrm{c}} \lesssim  \delta f \lesssim   1 / T_{\mathrm{OFDM}} \ll 100\mathrm{~kHz}.
    \label{eq:pnrange}
\end{equation}
With \eqref{eq:pllpn}, we compute the LO phase noises at these frequency offsets for the following devices described in Table~\ref{table:pn}.

\subsubsection{NI USRP B200}
We selected the NI USRP B200 as an example of a simple Software-Defined Radio. It features a locally thermal-compensated crystal oscillator (TCXO) reference that feeds into the Analog Devices AD9361 transceiver IC, which includes its own PLL and VCO.

% Variants:\\
% 2. Pluto aka B200 (check B200 shematics/manual): cots TXCO + AD9361 \footnote{Pluto manual}\footnote{B200 shematics/manual}

% \footnote{cots TCXO datasheet}
% \footnote{AD9361 datasheet with reference of particular nouse graphq}
% 1. USRP X300, LN TCXO + Clock Jitter Cleaner + dedicated LO\footnote{X300 shematics/manual}
% 3. Instrumental-grade? Check performance of, say, R\&S VSG and SigAn.
% 4. Virtual, perfect generator -- maybe.

\subsubsection{NI USRP RIO 2954R}

For a more advanced SDR, we consider the USRP RIO 2954R, which is based on the Ettus X300 series. This device incorporates a clock cleaning and distribution IC with two additional PLL loop stages: one utilizing a low-noise voltage-controlled crystal oscillator (LN-VCXO) and the other employing a high-frequency, low-noise VCO. The latter frequency can be divided as needed and is used for clocking the ADC/DAC and as a reference for the Maxim Integrated MAX2871 LO. For simplicity in our model, we utilize the free-running performance of the LN-VCXO as the reference XO and the PN of the MAX2871.

\subsubsection{The Instrumental Grade Transceiver}

As a reference point, we evaluate the performance of the Texas Instruments LMX2594 high-performance RF synthesizer IC, which is locked to the Symmetricom 1000C low-noise, high short-term stability oven-controlled crystal oscillator (OCXO).

\section{\label{sec:numres}Numerical Results}
%We assume that the Initial Access, Pilot Assignment, Cluster Formation, and radio resource allocation procedures are completed by utilizing techniques available in the literature \cite{beerten2023cellfree,Girycki}. 

%Consequently, we focus on assessing data transmission performance. 

%For the Frequency Range 1 (FR1: sub-6GHz), the most popular numerology is $\mu=0$, consequently, a TTI spans over 180 kHz bandwidth (12 subcarriers with 15 kHz spacing) and 1~ms (divided into 14 OFDM symbols with a duration of 71.4~$\mu$s each). In line with \cite{Girycki}, we assume the slot format where all symbols are used for uplink communication (format No. 1 defined in 3GPP 38.213 - Table 11.1.1-1).

In alignment with the selection of FR1, we consider  $M = 100$ APs equipped with $L = 4$ antennas and $K = 40$ UEs within $0.4 \times 0.4$ km$^2$ area.  Large-scale propagation effects are modeled as 
%\begin{equation}\label{eq:PL_log}
$\beta_{mk}(d)= \Lambda_0 + 10 \eta \log (d/d_0) + \Lambda_{sh},$
%\end{equation}
where $d$ is the distance between AP $m$ and UE $k$, $\Lambda_0=-35.3$ is the PL at reference distance $d_0=1$~m, pathloss exponent $\eta=3.76$, and $\Lambda_{sh}$ is lognormally distributed with the standard deviation of $\sigma_{sf}=10$. Moreover, UEs transmit with the same power $p = p_k = p_m = 100$~mW in both uplink training and transmission phases. The thermal noise variance is $\sigma_N$= -174~dBm/Hz. Regarding the PN model, we use the same parameters as in \cite{9528977,9502552} and \eqref{eq:var} for our time scales of one OFDM symbol $T_{\rm OFDM}$, the obtained variance of the innovation equals $0.23$ (note that \cite{Wu2023} used 4 times higher variance resulting in even more severe PN).

\begin{figure}
    \centering
    \includegraphics[width=0.85\columnwidth]{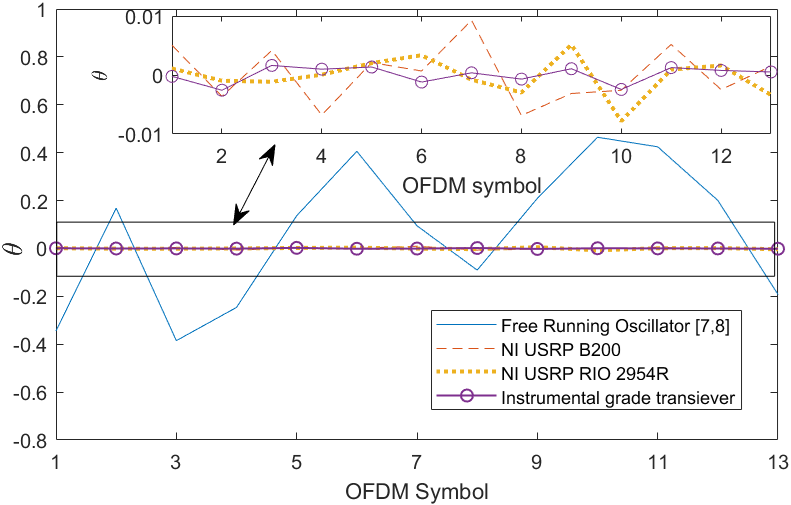}
    \caption{Comparison of the phase drift generated by different models: Realistic LOs are more stable.}
    \label{fig:PN comparison}
    \vspace*{-0.5cm}
\end{figure}

Fig.~\ref{fig:PN comparison} illustrates the differences in PN produced by the models used in \cite{9528977,9502552} compared to our study. The realistic LOs are significantly more stable than the free-running oscillator models employed in previous research on CF systems. The realistic LOs exhibit moderate drift, resulting in an order of magnitude smaller CPE than the earlier models \cite{9528977,9502552}. Moreover, ICI is negligible when using realistic LOs, whereas previous theoretical models generated PN that caused significant ICI.

\begin{figure}
\centerline{\includegraphics[width=0.85\columnwidth]{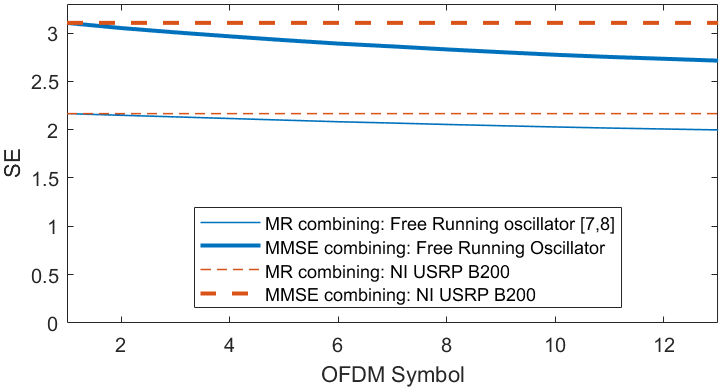}}
\caption{Average SE for different LOs: PN of the realistic LO does not cause performance degradation.}
\label{fig:avgSE}
\vspace*{-0.6cm}
\end{figure}

Fig.~\ref{fig:avgSE} depicts the variation in achievable SE per user across OFDM symbols due to the evolving CPE. Under severe PN conditions, as parameterized in models \cite{9528977,9502552}, the achievable SE decreases by {by 10\%} within one TTI due to the cumulative effect of PN drift. However, when employing the realistic PN model (based on the NI USRP B200, which has the least stable LO), we observe no degradation in SE within the TTI (1~ms).

\section{Conclusions}

This paper critically examines the performance of cell-free (CF) massive MIMO (mMIMO) systems under the influence of phase noise (PN) using a realistic hardware-based model. Unlike previous studies that relied on simplified oscillator models, our approach utilizes a more accurate representation of local oscillators (LOs) found in radios such as the NI USRP B200 and NI USRP RIO. While this paper focuses on CF mMIMO systems, the PN model can be applied to evaluate the performance of any RF system operating within the sub-6 GHz frequency range (FR1). The proposed OFDM CF mMIMO model with realistic PN serves as a valuable tool for assessing and designing practical CF systems, such as defining numerology and pilot allocation schemes. Our findings indicate that even less stable, cost-effective LOs maintain adequate phase stability to prevent Spectral Efficiency losses over the standardized 3GPP transmission time interval of 1 ms. These results affirm the feasibility of using existing 5G standards in the deployment of scalable CF mMIMO systems for future 6G infrastructure. The demonstrated resilience of CF mMIMO systems to hardware imperfections supports the practical implementation of these networks.
%, alleviating some of the concerns raised by previous theoretical studies. %As the industry moves towards practical deployments, our study underscores the importance of considering practical hardware limitations to ensure the robustness and efficiency of 6G networks.
\vspace*{-0.5cm}

%\section*{Acknowledgements}
%This research is supported by ?
%\as{Genia or Sofie to verify this}
\bibliographystyle{IEEEtran}
\bibliography{ref}

\end{document}